\documentclass[useAMS,usenatbib,referee]{mn2e}
\usepackage{graphicx}
\usepackage{amssymb}
\usepackage{natbib}
\title[WASP-1 Photometric follow-up]{Photometric follow-up of the transiting planet WASP-1b}
\author[Shporer et al.]{A. Shporer$^1$\thanks{E-mail: shporer@wise.tau.ac.il},
 O. Tamuz$^1$, S. Zucker$^2$ and T. Mazeh$^1$\\
  $^1$Wise Observatory, Raymond and Beverly Sackler Faculty of Exact Sciences, Tel Aviv University, \\
      Tel Aviv, Israel 69978\\
  $^2$Dept. of Geophysics \& Planetary Sciences, Raymond and Beverly Sackler Faculty of Exact Sciences, \\
      Tel Aviv University, Tel Aviv, Israel 69978}
\date{Released 2006 Xxxxx XX}

\pagerange{\pageref{firstpage}--\pageref{lastpage}} \pubyear{2006}

\def\LaTeX{L\kern-.36em\raise.3ex\hbox{a}\kern-.15em
    T\kern-.1667em\lower.7ex\hbox{E}\kern-.125emX}

\begin{document}

\label{firstpage}

\maketitle

\begin{abstract}
We report on photometric follow-up of the recently discovered
transiting planet WASP-1b. We observed two transits with the Wise
Observatory 1m telescope, and used a variant of the EBOP code
together with the Sys-Rem detrending approach to fit the light
curve. Assuming a stellar mass of $1.15 M_{\sun}$, we derived a
planetary radius of $R_p = 1.40\pm0.06 R_J$ and mass of $M_p =
0.87\pm0.07 M_J$. An uncertainty of $15\%$ in the stellar mass results
in an additional systematic uncertainty of $5\%$ in the planetary
radius and of $10\%$ in planetary mass.  Our observations yielded 
a slightly better ephemeris for the center of the transit: $T_c$ [HJD] $= (2454013.3127 \pm 0.0004)
+ N_{tr} \cdot (2.51996\pm0.00002)$.  The new planet is an inflated,
low-density planet, similar to HAT-P-1b and HD209458b.
\end{abstract}

\begin{keywords}
planetary systems -  stars: individual: WASP-1 - techniques: photometric
\end{keywords}

\section{Introduction}

Wide-field small-aperture telescopes are currently used by a few
groups (e.g., Bakos et al. 2004, Alonso et al. 2004, McCullough et
al. 2005, Pollacco et al. 2006) to search for transiting planetary
candidates. 
However, light curves obtained using wide-field
small-aperture telescopes are usually not accurate enough to put
useful constraints on the system parameters. Hence, photometric
follow-up using larger telescopes is essential.

The WASP\footnote{http://www.superwasp.org} consortium (Pollacco et
al. 2006) has recently detected two new transiting extrasolar planets,
WASP-1b and WASP-2b (Collier Cameron et al. 2006). The discovery paper
suggested that WASP-1b is probably an inflated, low-density planet,
similar to HAT-P-1b (Bakos et al. 2006) and HD209458b (e.g., Knutson
et al. 2006). 
However, using the photometry of the SuperWASP small-aperture cameras
and a single transit observed by a 35cm telescope, 
Collier Cameron et al. (2006) could not
constrain the system parameters of WASP-1 very well.  In this work we
have set out to better constrain these parameters using the Wise
Observatory 1m telescope. We describe our observations in \S~\ref{obs}
and the data processing in \S~\ref{proc}. In \S~\ref{sum} we briefly
discuss our results.

\section{Observations}
\label{obs}

We observed two WASP-1b transits with the 1m telescope at the Wise
Observatory, on the nights of 2006 October 4 and 2006 October 9.
The observations were carried out in the $I$ filter, with auto-guiding and no
defocusing, using a Tektronix $1024\times1024$ pixel
back-illuminated CCD, with a pixel scale of 0.696~arcsec~pixel$^{-1}$
and an $11.88~\times~11.88$ arcmin overall field of view (Kaspi et
al. 1999).

On 2006 October 4 the exposure time was $60$ seconds at the beginning
of the transit, at airmass $\sim2$, decreasing to $45$ seconds at
lower airmass. PSF FWHM was about $2.9$\,arcsec that night.  On 2006
October 9 exposure time varied between $20$ and $45$ seconds according
to varying observing conditions, in particular the bright moon's altitude.
PSF FWHM was about 2.1
arcsec that night. There was a short period of cloudiness on the
October 9 night which prevented us from observing the egress of that
transit. Nevertheless, we were able to observe the star for a short
period after the transit, thus allowing calibration of the light curve
zero point.

We did not center the field of view on the target star but instead 
positioned it at RA = 00h20m21s, DEC = +32$^\circ$02'39'' (J2000),
in order to include a maximum number of comparison stars, which are
essential for the transit light curve reduction.

\section{Data processing and model fitting}
\label{proc}

We used IRAF\footnote{IRAF (Image Reduction and Analysis Facility) is
distributed by the National Optical Astronomy Observatories (NOAO),
which are operated by the Association of Universities for Research in
Astronomy (AURA), Inc., under cooperative agreement with the National
Science Foundation.} \texttt{CCDPROC} package for the bias subtraction
and flat field correction, using calibration exposures taken
nightly. Using the IRAF \texttt{PHOT} task we applied aperture
photometry to all field stars in the reduced frames, with a few trial
values for the aperture radius and sky annulus size. For the 2006
October 4 frames we obtained the most satisfactory result with an
aperture of $7$\,arcsec and an annulus inner and outer radii of $35$
and $63$ arcsec, respectively.  For the 2006 October 9 frames, an
aperture radius of $6$\,arcsec and an annulus inner and outer radii of
$28$ and $56$\,arcsec, respectively, yielded the best result.

Following Winn, Holman and Roussanova (2006), we used nine reference
stars and normalized their flux light curves to unit median. We
combined these normalized light curves by a simple average and a
$3\sigma$ rejection, thus creating a normalized relative flux
comparison light curve.  We normalized the target star flux light
curve by dividing it by the comparison light curve. Finally, we fitted
a linear function to the out-of-transit measurements and divided the
light curve by this function.

\subsection{Photometric parameters}
  
We used the Eclipsing Binary Orbit Program (EBOP) code (Popper \&
Etzel 1981) together with the Sys-Rem (Tamuz, Mazeh \& Zucker
2005; Mazeh, Tamuz \& Zucker 2006) code in order to fit the transit
light curve to our photometric measurements.

EBOP is widely used for modeling eclipsing binary light curves, and
can be easily adapted to model transits (Gimenez 2006).  It does not
model proximity effects very well, but this is irrelevant for
transits.  EBOP consists of two modules: a light curve generator and a
differential corrections module. Following Tamuz, Mazeh \& North
(2006) we used only the light curve generator and applied our own
optimization program.

Sys-Rem is an algorithm designed to remove systematic effects from
photometric light curves without assuming any prior knowledge of the
effects. These effects may result from varying observing conditions
between observations, such as airmass and weather conditions. Each
systematic effect is a sequence of generalized ``airmasses'' 
$\{a_j; j=1,...,M\}$ assigned to each image, where the index
$j$ refers to the image number and $M$ is the number of
images. Sys-Rem estimates both the effect and a set of coefficients,
$\{c_i; i=1,...,N\}$, which are generalized ``colors'', assigned to 
each star $i$, where $N$ is the
number of stars. Essentially, Sys-Rem optimizes the effects and the
coefficients such that subtracting the product $c_ia_j$ from $r_{ij}$, the
$j$-th measurement of star $i$, will minimize the RMS of the set 
$\{r_{ij} - c_ia_j\}$ (Tamuz et al. 2005).

As a set of photometric measurements can include a few different effects,
in our analysis we first estimated $6$ systematic
effects for each night separately, using the nine reference stars. 
Then, we searched for a set of
$12$ generalized color coefficients, six for each of the two nights, 
together with six transit parameters, giving a total of $18$ parameters,
that would best fit the
target light curve. The six transit parameters included the mid-transit
time $T_c$, the stellar fractional radius $r_*=R_*/a$, where $a$ is the
orbital semi-major axis, the ratio of planetary radius to stellar
radius $k=R_p/R_*$, the impact parameter $b=\cos i\cdot a/(R_*+R_p)$
and two magnitude zero points for the two nights $I_1$ and $I_2$.

Transit light curves usually require two additional parameters -- the
period and a limb-darkening coefficient. We did not optimize for the
period and adopted the period published by Collier Cameron et
al. (2006). Using the temperature and gravity given by Collier Cameron
et al. (2006) we adopted a value of $u=0.37$ (Van Hamme 1993). When $u$ is left
as a free parameter, the best-fitting value is $u=0.21 \pm 0.08$, 
which we considered an unphysical result.

Given the $4$ parameters $T_c$, $r_*$, $k$ and $b$, all the other
best-fitting $14$ parameters can be solved for analytically. Thus we
can perform a simple grid search over a $4$-dimensional space for the 
minimum $\chi^2$.

After fitting the above model to the data, the residual RMS was found to be 
$1.8$ mmag in the first night and $2.4$ mmag in the second. We therefore set
the photometric errors of each measurement to be equal to the
corresponding RMS, and repeated the analysis.

\begin{figure}
\includegraphics[width=8cm,height=6cm]{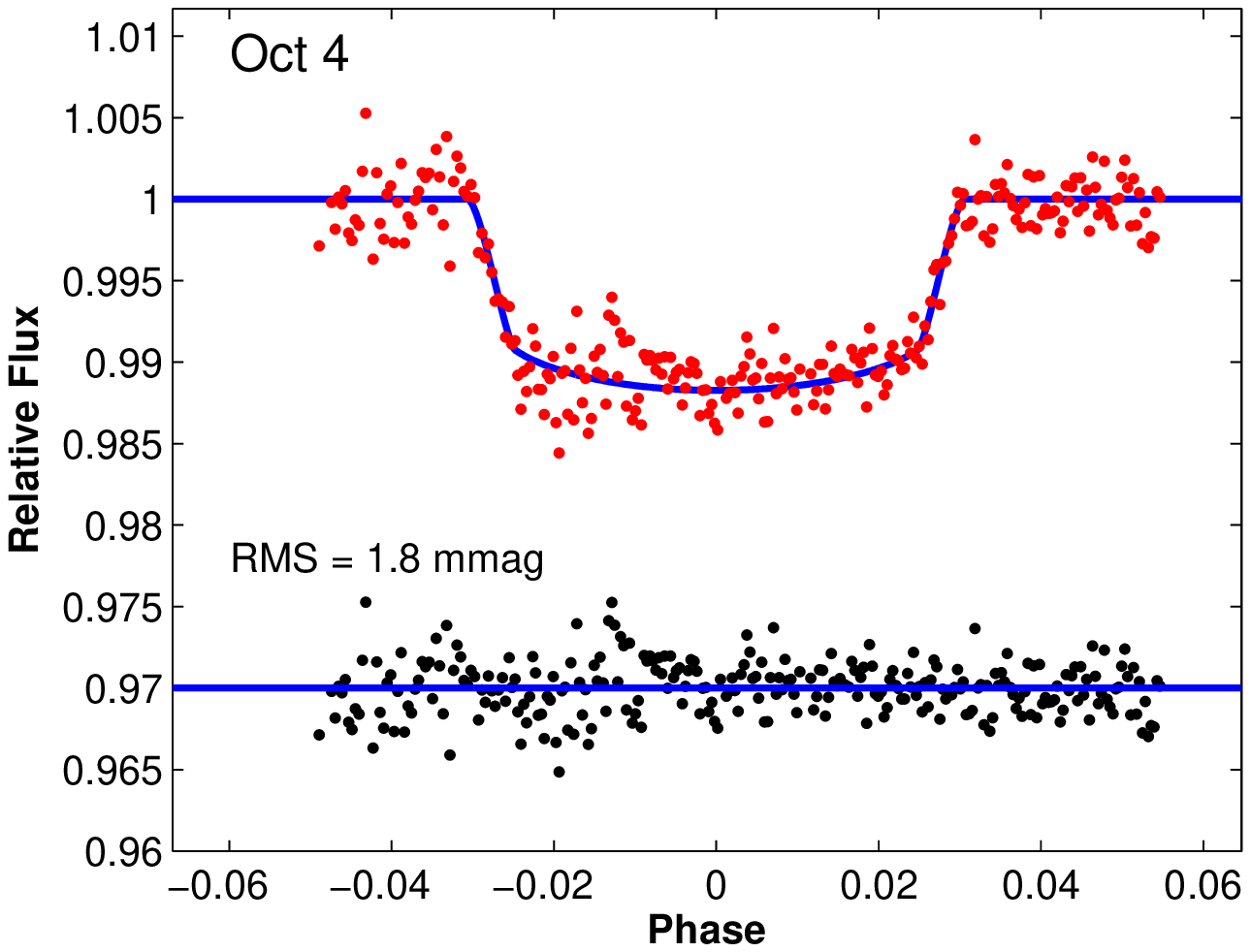}
\includegraphics[width=8cm,height=6cm]{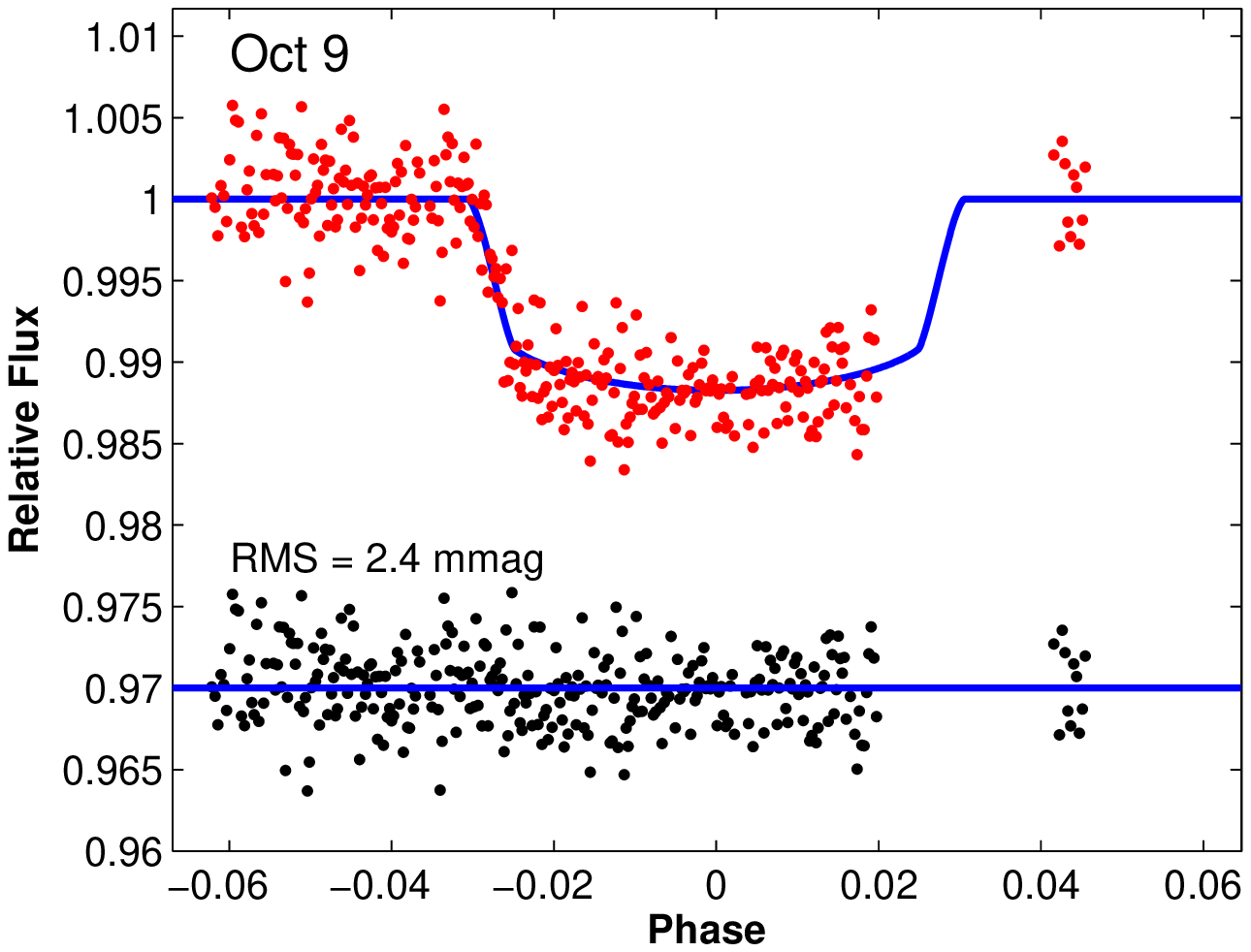}
\caption{Transit light curves of WASP-1, observed with the Wise Observatory
1m telescope on the nights of 2006 October 4 (left) and 2006 October 9 
(right). Relative flux is plotted against orbital phase and the best-fit 
model is overplotted. Residuals, derived by subtracting the model from the 
measurements, are plotted at the bottom of each panel and their
RMS in each of the light curves is also given.}
\label{lc}
\end{figure}

Table~\ref{lctab} lists all our Sys-Rem--detrended measurements.
 Table~\ref{params1} lists the
best-fitting values for each of the two transits independently and for
both transits simultaneously. 
We estimated the errors of the best-fitting values using
Monte Carlo simulations. We also give the best-fitting values derived
without using Sys-Rem, i.e., using only EBOP and no detrending.
Accuracy of these parameters is up to seven times better than
that of Collier Cameron et al. (2006). We adopted
values derived by fitting both nights simultaneously and applying 
Sys-Rem. Using these parameters we
estimate the orbital inclination to be $89.7 \pm 1.8$ deg.

\begin{figure}
\includegraphics[width=8cm,height=7cm]{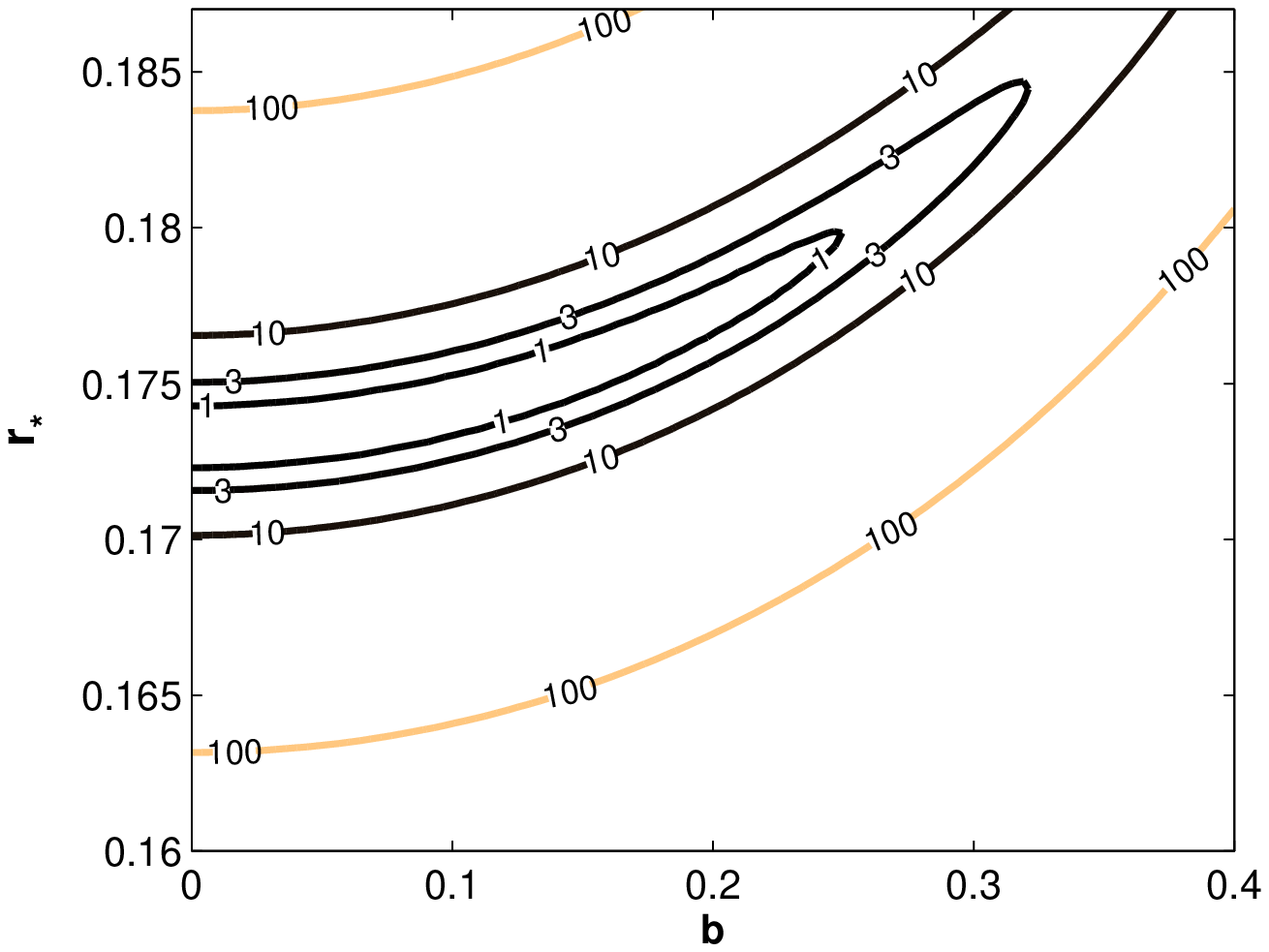}
\includegraphics[width=8cm,height=7cm]{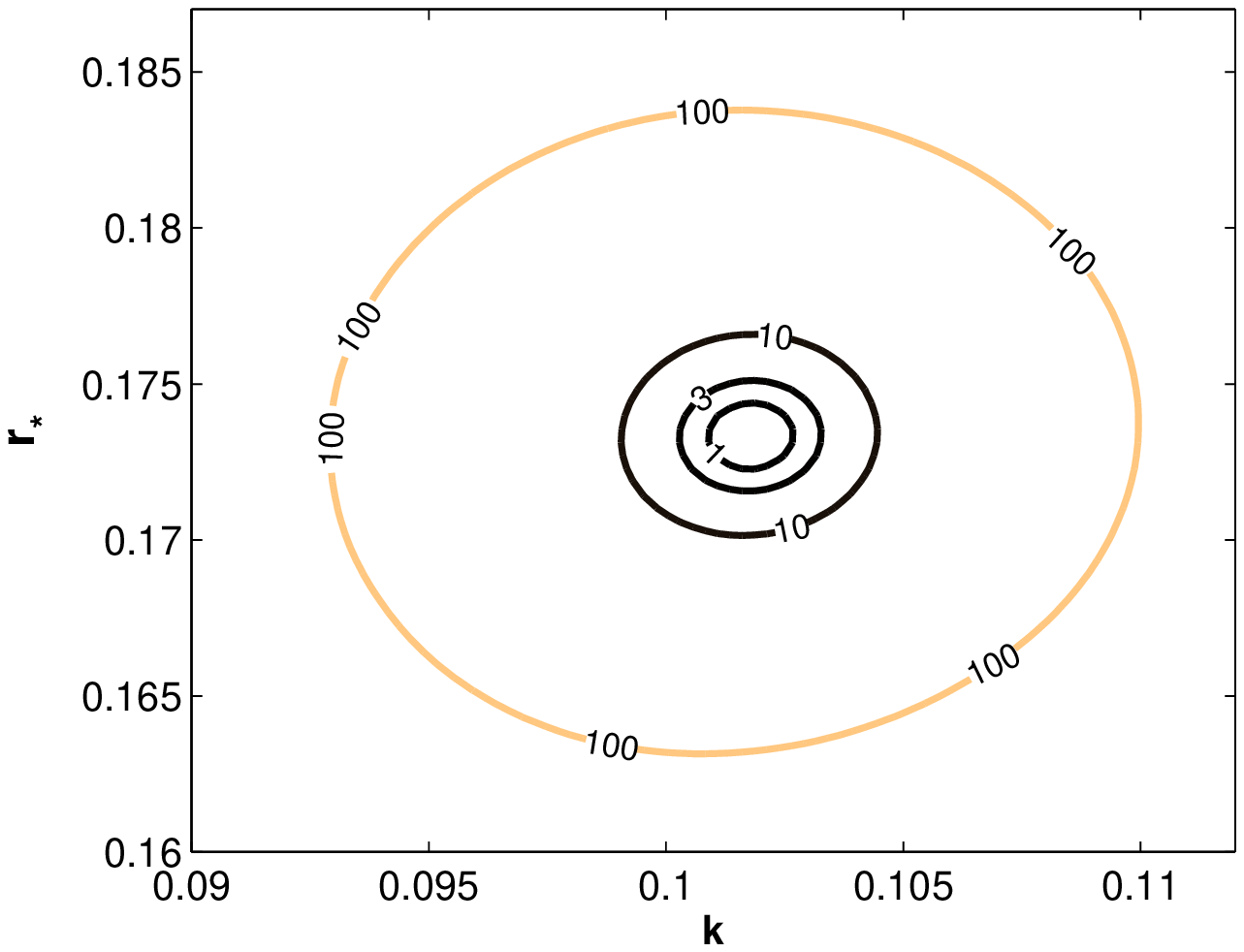}
\caption{Contour plots of the $\chi^2$ function, plotted vs. $r_*$ and $b$ (left) and $r_*$ and $k$ (right).
The contour labels denote the difference between the $\chi^2$ of that contour
and the minimal $\chi^2$.}
\label{contour}
\end{figure}

Fig.~\ref{lc} presents our light curves together with the fitted transit
model.
Fig.~\ref{contour} presents two $\chi^2$ contour maps, from
which we can learn about the relations among the best-fitting
parameters.  The left panel, presenting $\chi^2$ as a function of
$r_*$ and $b$, shows the well known degeneracy of the impact parameter
and the stellar radius, while the right panel, presenting $\chi^2$ as
a function of $r_*$ and $k$, shows that the two radii are practically
uncorrelated. 

\begin{table}
\caption{Table of all 583 photometric measurements. Only the first ten 
measurements are listed here, the complete table is available in the online 
version of this article in the Supplementary Materials Section.}
\label{lctab}
\begin{tabular}{ccc}
\hline
HJD - 2454000&  Rel. Flux& Err.\\
\hline
13.18941&  0.9971&  0.0016\\
13.19324&  0.9998&  0.0016\\
13.19432&  0.9982&  0.0016\\
13.19537&  1.0001&  0.0016\\
13.19642&  0.9997&  0.0016\\
13.19749&  1.0005&  0.0016\\
13.19854&  0.9979&  0.0016\\
13.19960&  0.9975&  0.0016\\
13.20065&  0.9987&  0.0016\\
13.20171&  0.9984&  0.0016\\
\vdots  &  \vdots &  \vdots\\ 
\hline
\end{tabular}
\end{table}

We used our value of mid-transit time, $T_c$, 
together with the one derived by Collier Cameron et al. (2006)\footnote{We
used a mid-transit time of HJD = $2453151.486 \pm 0.006$, from 2004, 
published in a preprint of Collier Cameron et al. (2006). 
This value was changed in the final version of the discovery paper, which was 
published only after we submitted this paper.} 
to re-calculate the orbital period, and obtained a value of 
$2.51996 \pm 0.00002$ days. There is no need to re-evaluate all the
other parameters, as the errors of the best-fitting parameters are not
dominated by the period error.

Our new transit elements allowed us to estimate the transit
total duration and duration of ingress and egress.  Adopting Eq.~4 of
Sackett (1999) for transit duration results in $t_T = 3.7\pm 0.2$
hours. Assuming transit ingress and egress are symmetric, duration of
ingress is calculated by:

\begin{equation}
\label{eq1}
t_{ingress} = \frac{P}{2\pi} \left[ \arcsin(\frac{\sqrt{(R_* + R_p)^2 - a^2\cos(i)^2}}{a}) - 
\arcsin(\frac{\sqrt{(R_* - R_p)^2 - a^2\cos(i)^2}}{a}) \right] ,
\end{equation}
which can be written also as:

\begin{equation}
t_{ingress} = \frac{P}{2\pi} \left[ \arcsin(r_*(1+k)\sqrt{1-b^2}) - \arcsin(r_*(1-k)\sqrt{1-b^2}) \right] .
\end{equation}
In our case $t_{ingress} = 20.7 \pm 0.9$ minutes.
  
\begin{table}
\caption{Light curve fitted parameters.}
\label{params1}
\begin{tabular}{lll}
                              & October 4 transit&\\
\hline
                              & Without Sys-Rem           & With Sys-Rem          \\
\hline
$r_* = R_* / a$               & $0.180 \pm 0.015$         & $0.179 \pm 0.019$      \\
$k = r_p/r_*$                 & $0.101 \pm 0.003$         & $0.100 \pm 0.003$      \\
$b = \cos i\cdot a/(R_*+R_p)$ & $0.24  \pm 0.19$          & $0.23  \pm 0.20$       \\
$T_c$ [HJD]                   & $2454013.3131 \pm 0.0006$ & $2454013.3114 \pm 0.0005$\\ 
\hline
\\
                              & October 9 transit&\\
\hline
                              & Without Sys-Rem           & With Sys-Rem          \\
\hline
$r_* = R_* / a$               & $0.187 \pm 0.016$         & $0.181 \pm 0.015$      \\
$k = r_p/r_*$                 & $0.107 \pm 0.002$         & $0.107 \pm 0.002$      \\
$b = \cos i\cdot a/(R_*+R_p)$ & $0.36 \pm 0.20$	          & $0.24 \pm 0.19$        \\
$T_c$ [HJD]                   & $2454013.3124 \pm 0.0045$ & $2454013.3114 \pm 0.0046$\\ 
\hline
\\
                              & Both transits&\\
\hline
                              & Without Sys-Rem           & With Sys-Rem          \\
\hline
$r_* = R_* / a$               & $0.186 \pm 0.015$         & $0.174 \pm 0.007$      \\
$k = r_p/r_*$                 & $0.104 \pm 0.002$         & $0.102 \pm 0.001$      \\
$b = \cos i\cdot a/(R_*+R_p)$ & $0.34 \pm 0.20$	          & $0.03 \pm 0.17$        \\
$T_c$ [HJD]                   & $2454013.3127 \pm 0.0005$ & $2454013.3127 \pm 0.0004$\\ 
\hline
\end{tabular}
\end{table}

\subsection{Radial-velocity elements}

Using the radial-velocity measurements supplied by Collier Cameron et
al. (2006) and our improved photometric elements, we were able to
re-calculate the radial-velocity elements $K_1$, the radial-velocity 
amplitude, and $\gamma$, the average radial-velocity. The
original SuperWASP photometric measurements were obtained two years before
the radial-velocities, and therefore did not constrain usefully their
orbital phases. Using our ephemeris, we were
able to determine the phases of the radial-velocities to a precision
of about a minute. Therefore, we have re-calculated the orbital parameters
using this constraint. Our derived radial elements are: $K_1 = 118 \pm
10$ m s$^{-1}$ and $\gamma = -13.506\pm0.008$ km s$^{-1}$.

In order to derive the orbital separation, stellar and planetary radii and the 
planetary mass we used the stellar mass given by Collier Cameron et al. (2006).
As there is a large uncertainty in the stellar mass, we repeated our 
calculation for the lower and upper limits of the published mass range. 
The results are given in Table~\ref{params2}. For the most likely value of the 
stellar mass, the planetary radius and mass are: 
$R_p = 1.40\pm0.06 R_J, M_p = 0.87\pm0.07M_J$, and the stellar radius is: 
$R_* = 1.42 \pm 0.06 R_{\sun}$. Since stellar and planetary radii scale 
as $M_*^{1/3}$ and the planetary mass scales as $M_*^{2/3}$, an 
uncertainty of 15\% on the stellar mass results in an additional systematic 
uncertainty of 5\% on the stellar and planetary radii and of 10\% in planetary 
mass. This systematic uncertainty should be added in quadrature to the errors 
in Table~\ref{params2}. 

\begin{table}
\caption{System parameters derived from the fitted parameters, assuming
three different values for the stellar mass. We use the most likely
value and the upper and lower limits of Collier Cameron et al. (2006).}
\label{params2}
\begin{tabular}{lllll}
\hline
$M_*$&        $a$&      $R_*$&            $R_p$&            $M_p$\\  
$[M_{\sun}]$& $[AU]$&   $[R_{\sun}]$&     $[R_J]$&          $[M_J]$\\
\hline
1.06&         0.037&  1.38$\pm$0.06&  1.36$\pm$0.06&  0.80$\pm$0.07\\
1.15&         0.038&  1.42$\pm$0.06&  1.40$\pm$0.06&  0.87$\pm$0.07\\
1.39&         0.041&  1.51$\pm$0.06&  1.49$\pm$0.06&  1.05$\pm$0.09\\
\hline
\end{tabular}
\end{table}

\section{Discussion}
\label{sum}

We present here photometry of two transits of the planet
WASP-1b, recently published by Collier Cameron et
al. (2006). Our new data better constrain the system parameters,
mainly the stellar and planetary radii. Combined with previously
published results, our new data provide a longer time span, which we
use in order to fix the phase of the radial-velocity orbit. 
Our analysis confirms Collier Cameron et al. (2006)
suggestion that the new planet is an inflated, low-density planet. 
We put all our photometric measurements in the public domain for any further
study.

In a simultaneous study, Charbonneau et al. (2006) conducted
follow-up observations of WASP-1 and WASP-2, and derived estimates 
for the stellar and planetary radii with the Mandel \& Agol
(2002) formulae. Although being slightly larger, their derived radii for the
WASP-1 system, of $R_* = 1.45 \pm 0.03 R_{\sun}$ and $R_p = 1.44 \pm
0.04 R_J$, are consistent with ours.

\begin{figure}
\includegraphics[width=8cm,height=7cm]{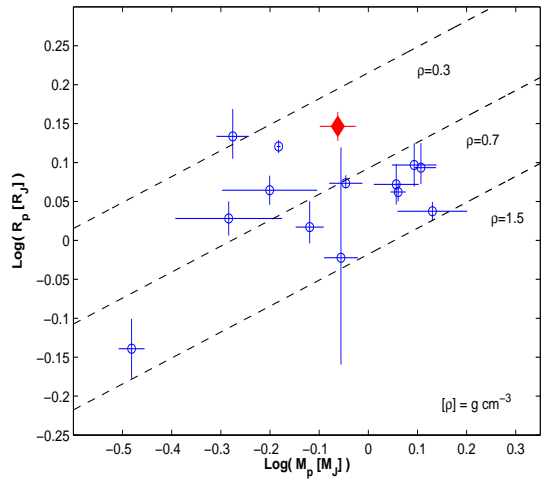}
\caption{Radius vs. mass of the 14 known transiting extrasolar planets in 
log-log scale. WASP-1b is marked by a filled diamond. Data for this 
figure was taken from http://exoplanet.eu/catalog-RV.php.}
\label{rm}
\end{figure}

Fig.~\ref{rm} presents the radii and masses of all currently known
transiting planets. The figure shows that WASP-1b radius is similar to
HAT-P-1b (Bakos et al. 2006) and HD209458b (e.g., Knutson et
al. 2006) radii. This small but growing group of inflated
extrasolar planets, whose radii is larger than predicted by
common theories of planet formation and evolution (e.g., Laughlin et
al. 2005), consists a significant fraction of all currently
known transiting planets. 
A number of possible explanations were suggested to account for this
discrepancy between theory and observations by considering, for example, 
an internal heat source as the cause of these large radii (e.g., Bodenheimer,
Laughlin \& Lin 2003, Winn \& Holman 2005).
However, in a recent paper, Burrows et al. (2006) suggest that enhanced 
planetary atmospheric metallicities, which increase atmospheric 
opacities, is the underlying mechanism responsible
for inflating these extrasolar planets. 
This explanation does not require any additional
heat source and is consistent with the increased probability of high
metallicity stars to host planets (Santos et al. 2005, 
Fischer \& Valenti 2005). In addition, Burrows et al. (2006) also pointed out 
that the commonly used analysis
of transiting light curves tend to derive radius larger than the one used in 
theory, which consider a planet radius till the point where the optical depth 
in the planet's atmosphere is $\tau = 2/3$. 

\begin{figure}
\includegraphics[width=8cm,height=7cm]{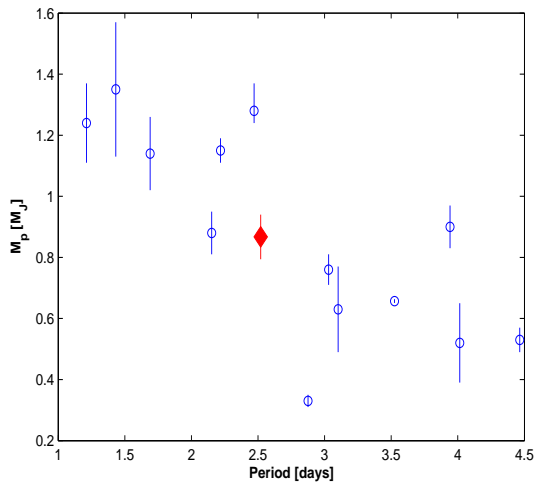}
\caption{Mass period relation for the 14 known transiting extrasolar planets.
WASP-1b is marked by a filled diamond. The planet with the lowest mass, 
at the bottom of the figure is HD149026b (Sato et al. 2005, 
Fortney et al. 2006). Data for this figure was taken from 
http://exoplanet.eu/catalog-RV.php.}
\label{massper}
\end{figure}

Combined with all the currently available mass and radius of
transiting extrasolar planets, our new values, plotted in
Fig.~\ref{massper}, are consistent with the mass-period relation
pointed out by Mazeh, Zucker \& Pont (2005) and by Gaudi, Seager \&
Mallen-Ornelas (2005). The only outlier to this relation is HD149026b
(Sato et al. 2005) which probably has a dense core (Fortney et
al. 2006).  We can gain a deeper understanding of this relation and
its possible origin by considering the energy diagram of Lecavelier
des Etangs (2006, his Fig.~1) presenting the surface potential energy of
{\it all} extrasolar planets versus the EUV energy flux they receive from
their host star. 
That diagram clearly shows a forbidden region, in which 
planets with masses too small evaporate because they absorb energy fluxes too 
large. 
Maybe the planets can not populate the left-hand bottom of our diagram
because of evaporation. However, this scenario does not account for the paucity
of transiting planets in the upper right of the diagram. Therefore, we have to
find many more transiting planets, in order to verify the mass-period relation 
and understand its nature. 

\section*{acknowledgments} 
We would like to thank Efi Hoory for his dedicated work while observing WASP-1
on the night of 2006 October 9. We also wish to thank Elia Leibowitz and 
Liliana Formiggini
for allowing us to use their telescope time on the night of 2006 October 4.  
We thank the anonymous referee for his thorough reading of the paper and his
comments which allowed us to improve this paper.
This work was supported by the Israeli Science Foundation through grant 
no. 03/323. This research has made use of NASA's Astrophysics Data System 
Abstract Service and of the SIMBAD database, operated at CDS, Strasbourg, 
France.



\label{lastpage}


\begin{thebibliography}{}

  \bibitem{} Alonso R., et al., 2004, AN, 325, 594

  \bibitem{} Bakos G., et al., 2004, PASP, 116, 266

  \bibitem{} Bakos G., et al., 2006, ApJ, In press (astro-ph/0609369) 

  \bibitem{} Bodenheimer P., Laughlin G., Lin D.~N.~C., 2003, ApJ, 592, 555 

  \bibitem{} Burrows A., Hubeny I., Budaj J., Hubbard W.~B., 2006, ApJ, submitted (astro-ph/0612703) 

  \bibitem{} Charbonneau D., et al., 2006, ApJ, submitted (astro-ph/0610589) 

  \bibitem{} Collier Cameron A., et al., 2006, MNRAS, In press (astro-ph/0603688) 

  \bibitem{} Fischer D.~A. \& Valenti J., 2005, ApJ, 622, 1102 

  \bibitem{} Gaudi B. S., Seager S., Mallen-Ornelas G., 2005, ApJ, 623, 472

  \bibitem{} Gimenez A., 2006, Ap\&SS, 304, 21

  \bibitem{} Kaspi S., Ibbetson P. A., Mashal E., Brosch N., 1999, Wise Observatory Technical Report 95/6

  \bibitem{} Knutson H., Charbonneau D., Noyes R. W.. Brown T. M., Gilliland R. L., 2006, ApJ, submitted (astro-ph/0603542)

  \bibitem{} Laughlin G., et al., 2005, ApJ, 621, 1072 

  \bibitem{} Lecavelier des Etangs A., 2006, A\&A, 461, 1185

  \bibitem{} Mazeh T., Zucker S., Pont F., 2005, MNRAS, 356, 955

  \bibitem{} Mazeh T., Tamuz O., Zucker S., 2006, astro-ph/0612418
 
  \bibitem{} McCullough P. R., et al., 2005, PASP, 117, 783

  \bibitem{} Pollacco D. L., et al., 2006, PASP, 118, 1407  

  \bibitem{} Popper D. M., Etzel P. B., 1981, AJ, 86, 102

  \bibitem{} Sackett P. D., 1999, in Mariotti, J. M., Alloin D., eds, NATO ASIC Proc. 532, Planets Outside the Solar System: Theory and Observations. Kluwer, Boston, p. 189

  \bibitem{} Santos N.~C., Israelian G., Mayor M., Bento J.~P., Almeida P.~C., Sousa S.~G., Ecuvillon A., 2005, A\&A, 437, 1127 

  \bibitem{} Tamuz O., Mazeh T., Zucker S., 2005, MNRAS, 356, 1466

  \bibitem{} Tamuz O., Mazeh T., North, P., 2006, MNRAS, 367, 1521

  \bibitem{} Van Hamme W., 1993, AJ, 106, 2096 

  \bibitem{} Winn J.~N. \& Holman M.~J., 2005, ApJ, 628, L159 

  \bibitem{} Winn J. N., Holman M. J., Roussanova A., 2006, ApJ, In press (astro-ph/0611404)

\end{thebibliography}
\end{document}